\documentclass[a4paper,11pt,final]{article}

\usepackage{amssymb,bm,times,graphicx}

\usepackage{hyperref}
\hypersetup{pdftitle={An extinction-survival-type phase transition in the probabilistic cellular automaton p182-q200}, pdfauthor={J. Ricardo G. Mendonça and Mário J. de Oliveira}, pdfsubject={Statistical Mechanics, Cellular Automata}, pdfkeywords={Probabilistic cellular automata, CA 182, CA 200, phase transition, directed percolation}, unicode=true, extension=pdf, pdfnewwindow=true, pdftoolbar=true, pdfmenubar=true, pdffitwindow=false, pdfstartview={FitH}, colorlinks=true, linkcolor=blue, citecolor=blue, urlcolor=blue}

\def\leq{\leqslant}
\def\geq{\geqslant}

\begin{document}

\pagestyle{empty}
\renewcommand{\thefootnote}{\fnsymbol{footnote}}

\begin{titlepage}

\begin{center}

{\Large \bf An extinction-survival-type phase transition in \\ 
            the probabilistic cellular automaton $p182$--$q200$}

\vspace{6ex}

{\large
{\bf J. Ricardo G. Mendon\c{c}a}\footnote{Corresponding author.}\footnote{Email: {\tt \href{mailto:jricardo@ime.usp.br}{\nolinkurl{jricardo@usp.br}}}}
{\bf and} {\bf M\'{a}rio J. de Oliveira}\footnote{Email: {\tt \href{mailto:oliveira@if.usp.br}{\nolinkurl{oliveira@if.usp.br}}}}
}

\vspace{1ex}

{\it {Instituto de F\'{\i}sica, Universidade de S\~{a}o Paulo} \\ \makebox[0pt][c]{Rua do Mat\~{a}o, Travessa~R 187, Cidade Universit\'{a}ria -- 05508-090 S\~{a}o Paulo, SP, Brazil}}

\vspace{6ex}

{\large \bf Abstract \\}

\vspace{2ex}

\parbox{120mm}
{We investigate the critical behaviour of a probabilistic mixture of cellular automata (CA) rules $182$ and $200$ (in Wolfram's enumeration scheme) by mean-field analysis and Monte Carlo simulations. We found that as we switch off one CA and switch on the other by the variation of the single paramenter of the model the probabilistic CA (PCA) goes through an extinction-survival-type phase transition, and the numerical data indicate that it belongs to the directed percolation universality class of critical behaviour. The PCA displays a characteristic stationary density profile and a slow, diffusive dynamics close to the pure CA~$200$ point that we discuss briefly. Remarks on an interesting related stochastic lattice gas are addressed in the conclusions.

\vspace{2ex}

{\noindent}{\bf Keywords}: Probabilistic cellular automata~$\cdot$ CA~182~$\cdot$ CA 200~$\cdot$ phase transition~$\cdot$ directed percolation

\vspace{2ex}

{\noindent}{\bf PACS 2010}: 05.70.Fh $\cdot$ 64.60.$-$i $\cdot$ 64.60.Ht

\vspace{2ex}

{\noindent}{\bf Journal ref}.: \href{http://dx.doi.org/10.1088/1751-8113/44/15/155001}{{\it J. Phys. A: Math. Theor.\/} {\bf 44}, 155001 (2011)}}

\end{center}

\end{titlepage}


\pagestyle{plain}

\renewcommand{\thefootnote}{\arabic{footnote}}
\setcounter{footnote}{0}

\section{Introduction}
\label{intro}

Cellular automata (CA, also for cellular automaton) have been a major theme of investigation since their introduction in the late 40's and early 50's of the last century as model systems for parallel processing machines and algorithms, memory storage devices, and self-reproducing simple organisms \cite{jvn}. Meanwhile, CA have become deeply intertwined with the subject matter of modern equilibrium and nonequilibrium statistical mechanics, such as with the theory of phase transitions, irreversibility, ergodicity, chaos, percolation, and physical kinetics, to name a few \cite{discrete,wolfram,dkpca,grinstein,rujan,lebowitz,chen,helbing,droz}.

Probabilistic versions of CA (PCA) with absorbing states, {\it i.e.\/}, with states that once reached trap the dynamics definitely, are a whole chapter in CA theory. The single most representative PCA in this class is the Domany-Kinzel (DK) PCA, because its two-dimensional parameter space encompasses both the directed bond and the directed site percolation processes \cite{dkpca}. More general PCA with more than one absorbing state have also been investigated, with larger neighbourhoods and more than two parameters, at the expense of some symmetries present in the DK PCA. In particular, a class of two and three parameter PCA with two absorbing states was investigated in \cite{boccara}, where many general results on the phase diagrams of the models were obtained.

In usual PCA modeling, the freedom to set the transition probabilities directly in the rule table allows one to embody competing interactions in the PCA from the outset, such that a resulting complex dynamic behaviour becomes an expected treat. However, elementary one-dimensional CA are more charming in their simplicity, since most of them can be implemented with a few deterministic binary operations that also make them of greater technological relevance.\footnote{An ``atlas'' of elementary CA can be consulted at {\tt http://atlas.wolfram.com/01/01/}.} Moreover, it has been found that by composing simple CA in space and time, probabilistically or not, we may obtain complex behaviour out of simple components. A striking example of this possibility is given by the composite deterministic CA of H.~Fuk\'{s} that solves the density classification problem, an impossible problem for locally interacting single CA \cite{fuks,noisy}. These observations triggered our interest in probabilistic mixtures of simple CA. One PCA that we considered preliminarly combines CA rules $150$ and $200$ in Wolfram's enumeration scheme \cite{wolfram} and is briefly mentioned in Section~\ref{mcarlo}. Another possible combination is given by CA rules $23$ and $254$, that when combined as a PCA in the guise of a model for an organism ($1$'s) consuming finite renewable resources ($0$'s) and dying from overcrowding may display an active-inactive type phase transition. As far as we are aware, this kind of composite PCA has not yet been fully explored in the literature.

In this paper we investigate in detail a probabilistic mixture of CA~$182$ and CA~$200$ that is left-right symmetric, has a single scalar order paramenter, no conserved quantities, and two absorbing states, both reachable dynamically as long as the mixed PCA contains finite portions of both CA~$182$ and CA~$200$ dynamics. After some exploratory work, we found that this PCA displays some interesting features, such as its behaviour in a certain small parameter region besides an extinction-survival-type phase transition as we probabilistically switch off one CA and switch on the other. We characterize this phase transition by mean-field analysis and Monte Carlo simulations and determine that it belongs to the directed percolation universality class of critical behaviour.

The article is organized as follows. In Section~\ref{model}, we define the PCA and in Section~\ref{mean} we analyse it at the mean-field level of approximation. Section~\ref{mcarlo} presents the results of direct Monte Carlo simulations of the PCA for its critical point and critical exponents, and in Section~\ref{summary} we make a few final remarks, summarize our results, and identify some perspectives for further investigation.


\section{Model description}
\label{model}

Let $\eta_{\ell}(t) \in \{0,1\}$ denote the state of the site $\ell \in \Lambda\subset {\mathbb Z}$ at instant $t \in {\mathbb N}$, with $\Lambda$ a finite lattice of $L$ sites with periodic boundary conditions $\ell+L \equiv \ell$. The state of the system at instant $t$ is given by $\bm{\eta}(t) = (\eta_{1}(t), \eta_{2}(t), \ldots, \eta_{L}(t)) \in \{0,1\}^{\Lambda}$. The model we are interested in is the probabilistic cellular automaton defined by the rules in Table \ref{TAB-W}. We dub this system PCA $p182$--$q200$, with $p+q=1$.

When $p=1$, the mixed PCA reduces to the deterministic CA~$182$, which has a stationary density of 1's given by $\rho_{182}^{\rm exact} = 3/4$ \cite{wolfram}. CA~$200$ ($p=0$ in Table~\ref{TAB-W}), otherwise, is quite an uninteresting CA, since most initial configurations die out quickly or do not evolve at all under its rules. The role of CA~$200$ in our mixture of CAs is to provide a route to the absorbing state devoid of 1's, a state that cannot be reached by CA~$182$ except from the initial empty configuration itself. So, as $p$ varies between $p=0$ and $p=1$ we expect a competition to set up between the two rules to dominate the dynamics, with an extinction-survival-type phase transition somewhere in between. We indeed found such a phase transition, as long as the initial condition is not one of the absorbing configurations of the model.

PCA $p182$--$q200$ can be related with the DK PCA only at $p = 1/2$, a point that is off the mixed site-bond percolation parameter subspace of the DK PCA, since this subspace requires that a certain parameter $x$ of the DK PCA be zero, which is impossible in our setting, because the match between the two PCAs requires that $p = 1/2$ and $x=1-p$ simultaneously. The point $p=1/2$ in PCA $p182$--$q200$, however, corresponds to a completely uncorrelated dynamics, since bits will flip with probability $1/2$ irrespective of their neighborhood, with the exception of the bits in the bulk of blocks $11{\cdots}1$, {\it cf.\/}~Table~\ref{TAB-W}. These blocks, however, are not stable at $p \neq 0$, since they will eventually be eroded from the boundaries. We thus expect that at $p=1/2$ the stationary density of active sites fluctuates around $\rho = 1/2$.

\begin{table}[t]
\centering
\caption{\label{TAB-W}\small Rule table for PCA $p182$--$q200$, $p+q=1$. The first row gives the initial neighborhood, the other two rows give the final state reached by the central bit of the initial neighborhood with the probability given at the leftmost column. Clearly, the configurations $00{\cdots}0$ and $11{\cdots}1$ are absorbing configurations of the PCA.}
\medskip
\begin{tabular}{ccccccccc}
\hline \hline
        & 111 &  110  &  101  & 100 &  011  & 010 & 001 &  000 \\ \hline
$p$     &  1  &   0   &   1   &  1  &   0   &  1  &  1  &   0  \\
$q$     &  1  &   1   &   0   &  0  &   1   &  0  &  0  &   0  \\ \hline \hline
\end{tabular}
\end{table}


\section{Mean-field analysis}
\label{mean}

We begin by analysing PCA $p182$--$q200$ in the mean-field approximation to obtain some first information on its critical behaviour. For a nice brief exposition and application of the technique see, {\it e.g.\/}, \cite{tania}.

The dynamics of the probability distribution $P_{t}(\bm{\eta})$ of the states $\bm{\eta}$ of the PCA is ruled by the equation
\begin{equation}
\label{PPP}
P_{t+1}(\bm{\eta}') = \sum_{\bm{\eta}}W(\bm{\eta}'|\bm{\eta})P_{t}(\bm{\eta}),
\end{equation}
where the summation runs over all $\bm{\eta} \in \{0,1\}^{\Lambda}$ and $W(\bm{\eta}'|\bm{\eta}) \geq 0$ is the conditional probability for a transition $\bm{\eta}$ $\to$ $\bm{\eta}'$ to occur in one time step. Since in a CA or PCA all sites are updated simultaneously and independently, we can write
\begin{equation}
\label{WWW}
W(\bm{\eta}'|\bm{\eta}) =
\prod_{\ell=1}^{L}W_{\ell}({\eta_{\ell}'}|\bm{\eta}), \quad {\rm with} \quad
\sum_{\eta_{\ell}'}W_{\ell}({\eta_{\ell}'}|\bm{\eta}) = 1.
\end{equation}
For PCA $p182$--$q200$ we have $W_{\ell}({\eta_{\ell}'}|\bm{\eta})=$ $W({\eta_{\ell}'}|\eta_{\ell-1}, \eta_{\ell}, \eta_{\ell+1})$, independent of $\ell$. The time evolution of the marginal probability distribution $P_{t}(\eta_{\ell},$ $\eta_{\ell+1},$ \ldots, $\eta_{\ell+n-1})$ of observing $n$ consecutive sites in state $(\eta_{\ell}, \eta_{\ell+1}, \ldots, \eta_{\ell+n-1})$ is, from Eqs.~(\ref{PPP}) and (\ref{WWW}), given by
\begin{eqnarray}
P_{t+1}(\eta_{\ell}', \eta_{\ell+1}', \ldots, \eta_{\ell+n-1}') = \nonumber \\
= \sum_{\eta_{\ell-1}} \sum_{\eta_{\ell}} \cdots \sum_{\eta_{\ell+n}} W(\eta_{\ell}'|\eta_{\ell-1},\eta_{\ell},\eta_{\ell+1})\, W(\eta_{\ell+1}'|\eta_{\ell},\eta_{\ell+1},\eta_{\ell+2}) \ldots \nonumber \\
\ldots W(\eta_{\ell+n-1}'|\eta_{\ell+n-2},\eta_{\ell+n-1},\eta_{\ell+n})\,
P_{t}(\eta_{\ell-1}, \eta_{\ell}, \ldots, \eta_{\ell+n}).
\label{MARG}
\end{eqnarray}

We see from equation~(\ref{MARG}) that to determine the probability of observing $n$ consecutive sites in a given state at instant $t+1$ we need to know the probabilities of observing the state of $n+2$ sites at instant $t$. To proceed with the calculations in an approximate fashion, we truncate this hierarchy at some point to split the correlations and get a closed set of equations. The simplest approximation ($n=1$) is obtained by taking
\begin{equation}
\label{APPROX}
P_{t}(\eta_{\ell-1},\eta_{\ell},\eta_{\ell+1}) \approx P_{t}(\eta_{\ell-1})\, P_{t}(\eta_{\ell})\, P_{t}(\eta_{\ell+1}).
\end{equation}
Higher order approximations ($n \geq 2$) are obtained (assuming spatial homogeneity) by the generalized splitting scheme
\begin{equation}
\label{MFN}
P_t(\eta_{\ell-1}, \ldots, \eta_{\ell+n}) \approx
\frac{P_t(\eta_{\ell-1}, \ldots, \eta_{\ell+n-2})\, P_t(\eta_{\ell}, \ldots, \eta_{\ell+n-1})\, P_t(\eta_{\ell+1}, \ldots, \eta_{\ell+n})}
{P_t(\eta_{\ell}, \ldots, \eta_{\ell+n-2})\,P_t(\eta_{\ell+1}, \ldots, \eta_{\ell+n-1})}.
\end{equation}
From the rules in Table~\ref{TAB-W} and the above equations, the single-site ($n=1$) approximation for $\rho^{(1)}_{t} = P_{t}(\eta_{\ell}=1)$ (where the superscript refers to the order of the approximation) reads
\begin{equation}
\rho^{(1)}_{t+1} = (\rho^{(1)}_{t})^{3} + (2-p)(\rho^{(1)}_{t})^{2}(1-\rho^{(1)}_{t}) + 3p\rho^{(1)}_{t}(1-\rho^{(1)}_{t})^{2}. 
\end{equation}
In the stationary state, $\rho^{(1)}_{t+1} = \rho^{(1)}_{t} = \rho^{(1)}$, and the above equation becomes
\begin{equation}
\label{STAT}
\rho^{(1)} = (\rho^{(1)})^{3} + (2-p)(\rho^{(1)})^{2}(1-\rho^{(1)}) + 3p\rho^{(1)}(1-\rho^{(1)})^{2},
\end{equation}
with solutions $\rho^{(1)}=0$, $\rho^{(1)}=1$ and $\rho^{(1)}=(3p-1)/(4p-1)$. The first two solutions correspond to the two absorbing states of the dynamics, whereas the last, nontrivial solution corresponds to the active state and is valid as long as $p \geq 1/3$. The single-site mean-field approximation for PCA $p182$--$q200$ thus predicts an extinction-survival-type phase transition at $p_{c}^{(1)} = 1/3$. Notice that at $p=1$, $\rho^{(1)}=2/3$, not far from the exact stationary value $\rho_{182}^{\rm exact} = 3/4$ for CA~$182$.

\begin{figure}
\centering
\includegraphics[viewport = 64 54 534 798, scale=0.35, angle=-90]{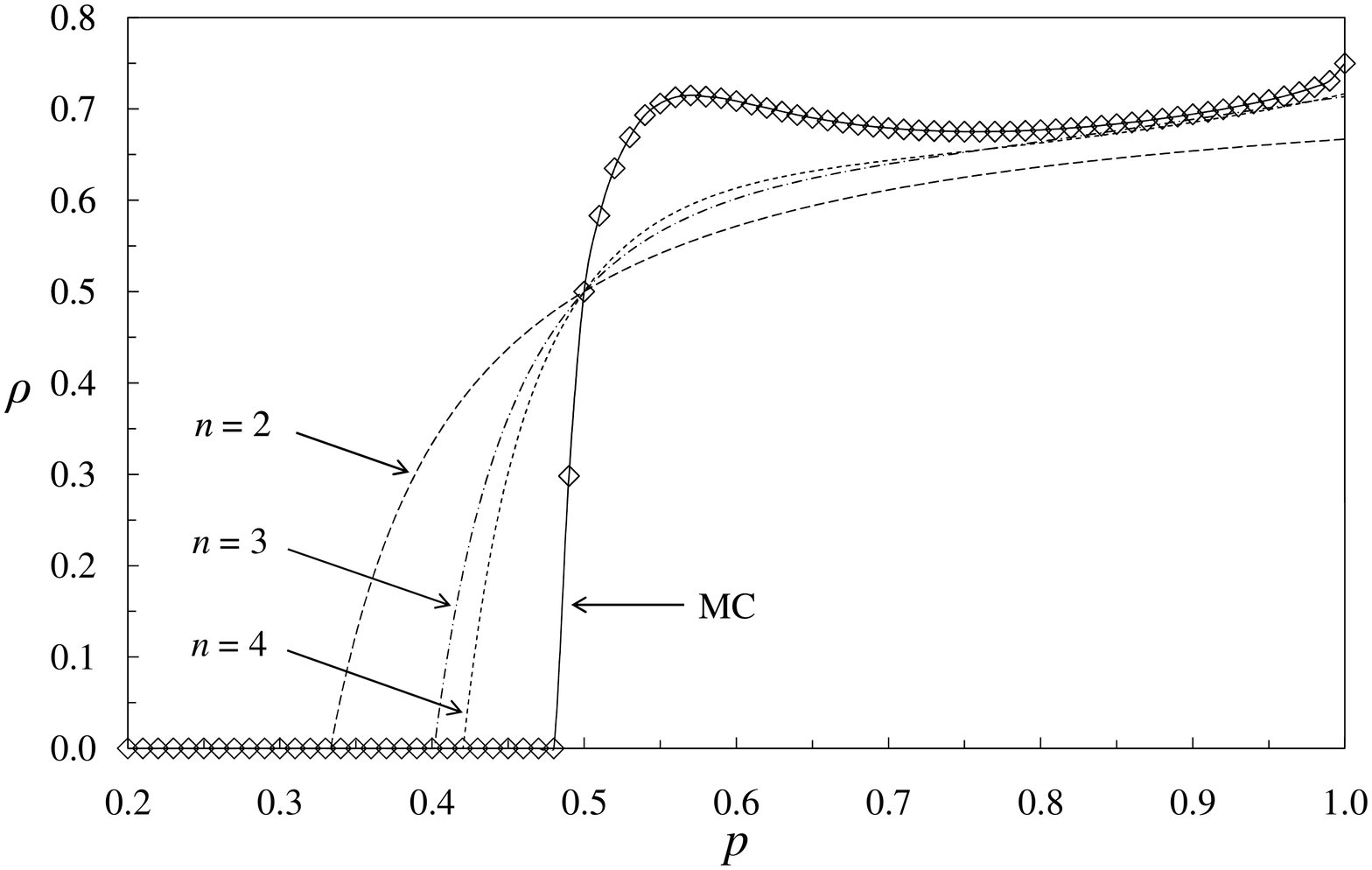}
\caption{\label{fig:pc}\small Density $\rho^{(n)}(p)$ of active sites in the mean-field approximations of orders $n=2$, $3$, and $4$, together with Monte Carlo simulation data for a lattice of $L=8000$ sites initialized randomly with density $1/2$. Each symbol in the Monte Carlo curve is an average over $10^{6}$ samples. Errors in the data are negligible below the critical point $p_{c} \simeq 0.48$, of the order of $5\%$ just at the critical point, and less than $0.2\%$ on the rest of the curve.}
\end{figure}

We have also considered higher-order approximations with $n=2$, $3$, and $4$. In these cases, however, the equations are too cumbersome to be written down here. The $n=2$ approximation was solved analytically and gives the same results as the $n=1$ case---somehow, the two-sites marginal probability $P(\eta_{\ell},\eta_{\ell+1})$ factors into $P(\eta_{\ell})P(\eta_{\ell+1})$. The cases $n=3$ and $4$ were solved numerically. The densities $\rho^{(n)}(p)$ of active sites for these approximation are shown in Figure~\ref{fig:pc}. As expected, at $p = 1/2$ all approximations (as well as the Monte Carlo simulation results, {\it cf.\/}~Section~\ref{mcarlo}) give $\rho(1/2) = 1/2$. In Table~\ref{TAB-PC} we show the values of $p_c^{(n)}$ and $\rho^{(n)}$ at $p=1$ for these approximations together with the corresponding values obtained by Monte Carlo simulations. As one can see from Table~\ref{TAB-PC}, the $n$-th order mean-field approximation to the exact PCA converges slowly with~$n$.

\begin{table}[t]
\centering
\caption{\label{TAB-PC}\small Critical parameter $p_{c}^{(n)}$ and value of the density of active sites $\rho^{(n)}(p)$ at $p=1$ obtained by mean-field approximations or order $n$ and Monte Carlo simulation ({\it cf.\/}~Section~\ref{mcarlo}). The numbers between parentheses (in this table and elsewhere in this article) indicate the uncertainty in the last digit(s) of the data. For $p=1$, PCA $p182$--$q200$ reduces to CA~$182$, for which the exact stationary density $\rho_{182}^{\rm exact} = 3/4$.}
\medskip
\begin{tabular}{cccc}
\hline \hline
$n$   & $p_{c}^{(n)}$ & $\rho^{(n)}(p=1)$ \\
\hline
1     &   1/3      &   2/3     \\
2     &   1/3      &   2/3     \\
3     & 0.4015     & 0.7135    \\
4     & 0.4203     & 0.7166    \\
MC    & 0.48810(5) & 0.7500(5) \\
Exact &    NA   &   3/4     \\
\hline \hline
\end{tabular}
\end{table}


\section{Direct Monte Carlo simulation}
\label{mcarlo}

\subsection{The density profile}

Our Monte Carlo simulations of PCA $p182$--$q200$ ran as follows. For each $p$, the PCA is initialized according to a Binomial\,($L$,\,$s$) distribution (mostly with $s=\frac{1}{2}$), {\it i.e.\/}, each site is initialized as $\eta_{\ell}=1$ with probability $s$, and relaxed through $L^{2}/10$ Monte Carlo steps, with one Monte Carlo step equal to a synchronous update of the states of all $L$ sites of the lattice. We then sample $\rho_{L} = L^{-1}\sum_{\ell}\eta_{\ell}$ every other Monte Carlo step.

\begin{figure}
\centering
\includegraphics[viewport= 72 49 519 795, scale=0.35, angle=-90]{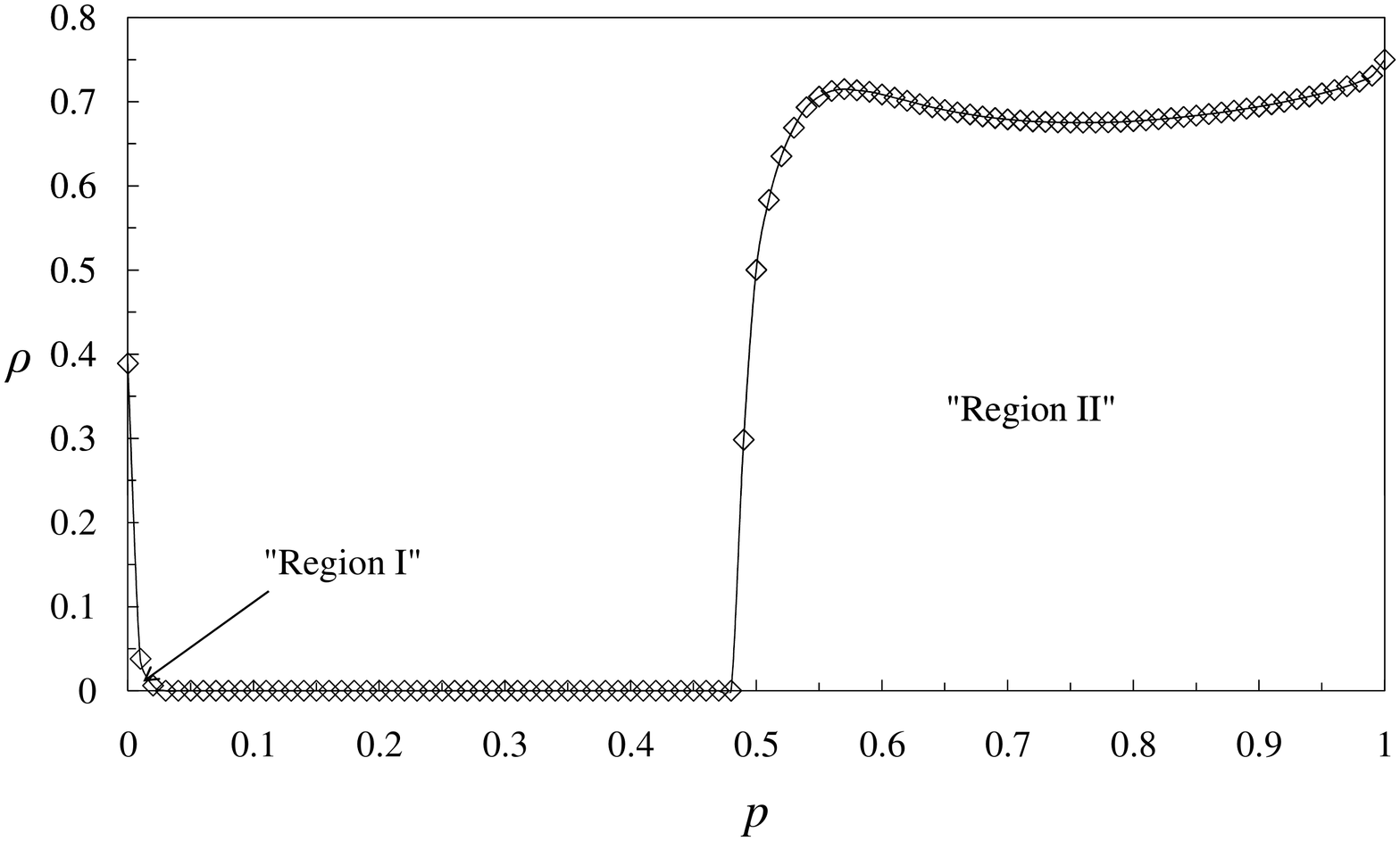}
\caption{\label{fig:rho}\small Density $\rho_{L}(p)$ of active sites for a lattice of $L=8000$ sites initialized randomly with density $1/2$. Each value of $\rho_{L}(p)$ is an average over $10^{6}$ samples after relaxation through $\sim L^{2}$ MCS. Errors in the Monte Carlo data are similar to the ones reported in Figure~\ref{fig:pc}.}
\end{figure}

Our data for $\rho_{L}(p)$ in a lattice of $L=8000$ sites appear in Figure \ref{fig:rho}. We can identify two regions in this figure: an ``active region I'' that goes from $p=0$ up to $p \simeq 0.03$ and an ``active region II'' that goes from $p \simeq 0.48$ up to $p=1$. Region I is easily understood as an artifact of our initialization of the PCA with a Binomial\,($L$,\,$\frac{1}{2}$) distribution. At $p=0$, PCA $p182$--$q200$ becomes CA~$200$, that washes out all local configurations but those of neighbouring 1's ({\it cf.\/}~Table~\ref{TAB-W}). The only neighborhood that evolves under CA~$200$ dynamics is 010, that becomes 000 with probability $q=1-p=1$. For a Binomial\,($L$,\,$s$) initial distribution, the expected initial density of 1's is $\mathbb{E}[\eta_{\ell}=1] = s$ and of triplets 010 is $\mathbb{E}[\eta_{\ell}\eta_{\ell+1}\eta_{\ell+2}=010] = (1-s)s(1-s)$. After one single time step, all 010 go into 000 and the stationary density of 1's becomes $\mathbb{E}[\eta_{\ell}=1] = s-(1-s)s(1-s)$. For $s=1/2$, $\rho(p=0) = 3/8$, in accordance with the value in Figure~\ref{fig:rho}. As $p$ increases from zero, the noise brough up by the CA~$182$ rules disturbs the dynamics and the PCA eventually converges to the absorbing state devoid of active sites.

\begin{figure}
\centering
\begin{tabular}{cc}
\includegraphics[viewport= 70 129 510 718, scale=0.25, angle=-90]{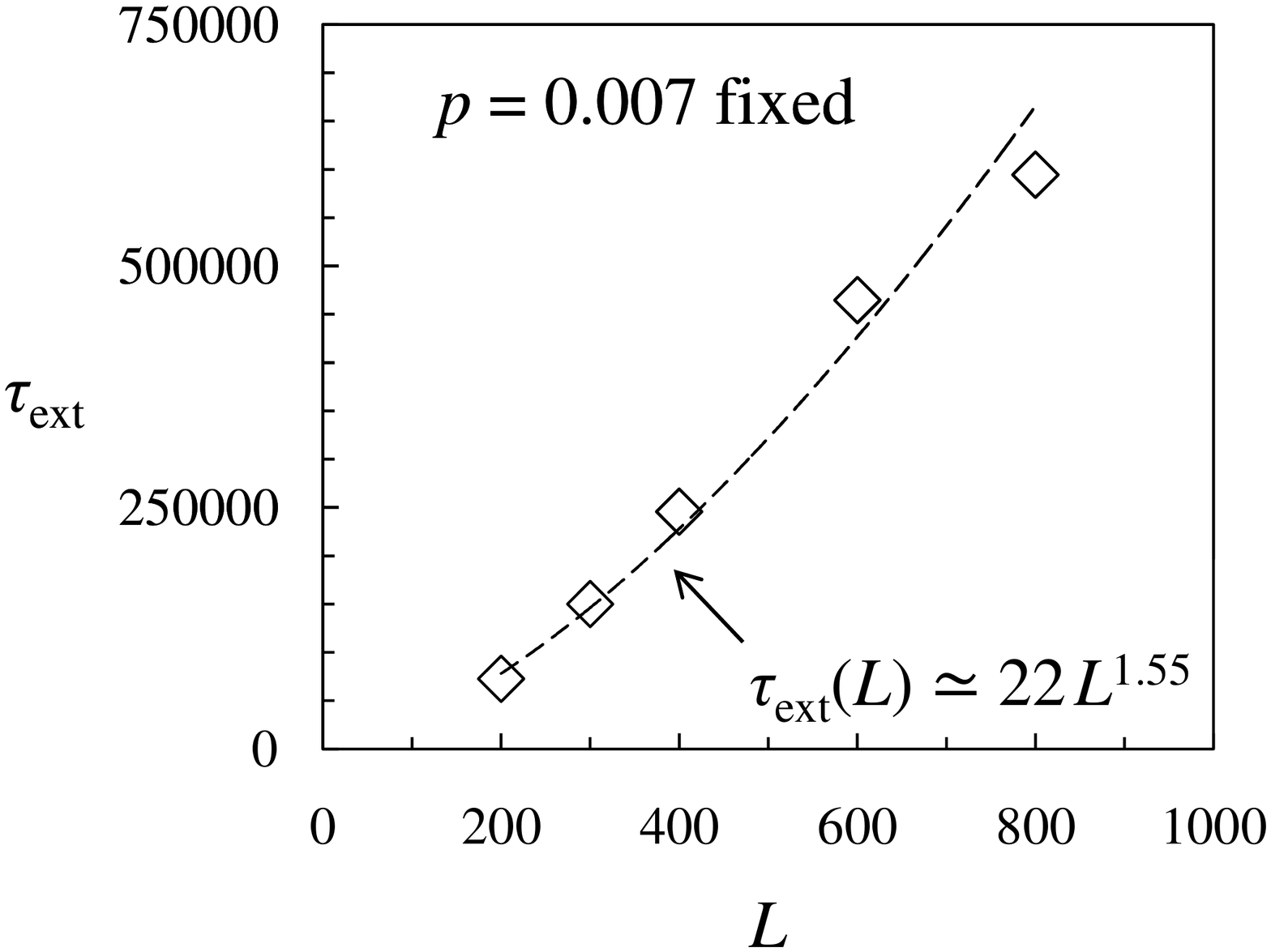} &
\includegraphics[viewport= 70 130 517 735, scale=0.25, angle=-90]{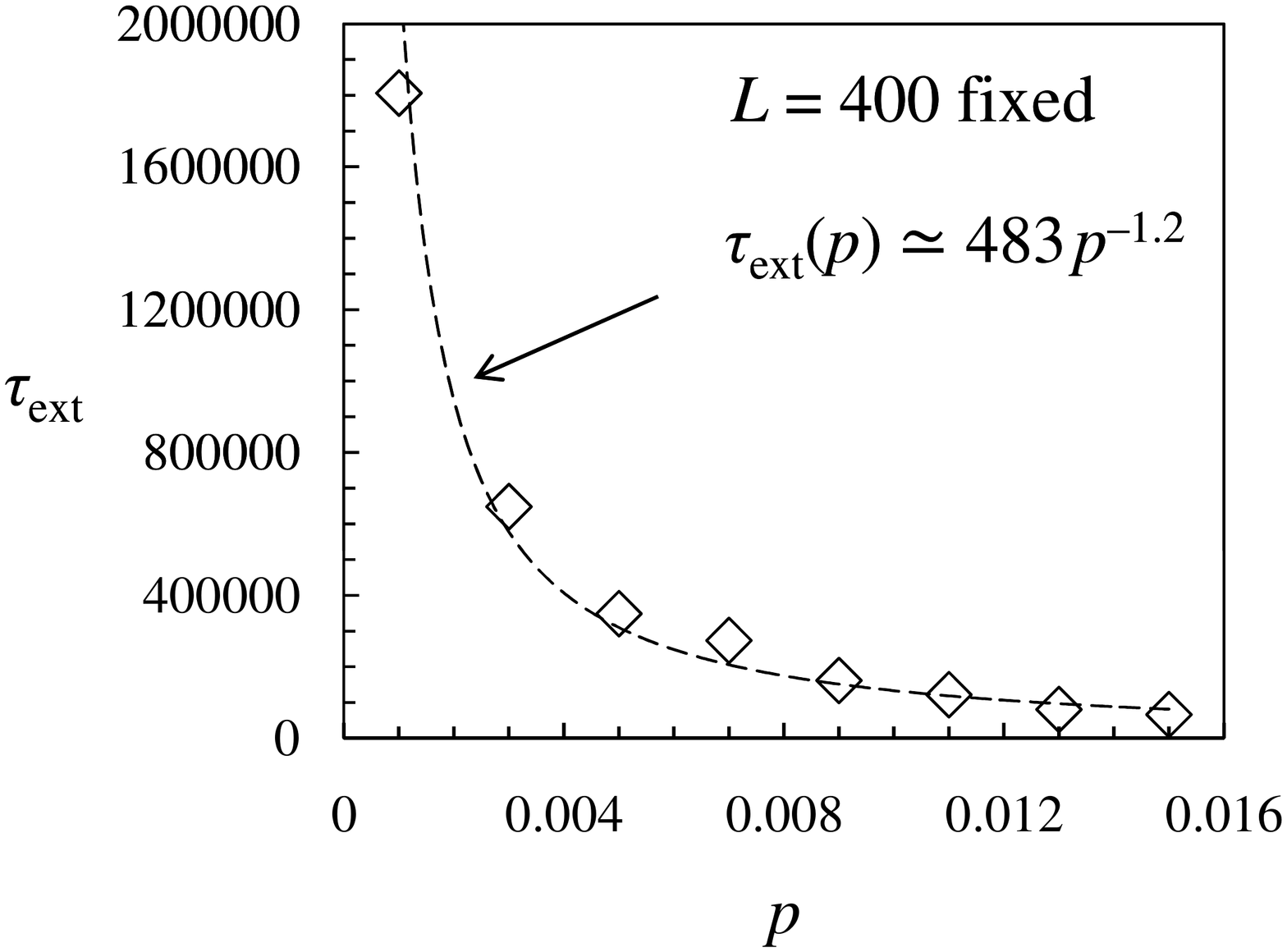}
\end{tabular}
\caption{\label{fig:tau}\small Extinction time $\tau_{\rm ext}(L,p)$ for small $p$. Each point shown is an average over 1000 realizations of the PCA with a random initial distribution of active sites with density $\rho(0)=0.3$. The curves fitted are guides only---the prefactors vary widely since only a tiny fraction of all possible initial conditions are sampled.}
\end{figure}

For very small $p > 0$, after the triplets $010$ quickly become $000$ with probability $q=1-p$, only the boundaries of the remaining clusters $11{\cdots}1$ move, performing random walks $100 \rightleftharpoons 110$ at the right edge and $001 \rightleftharpoons 011$ at the left edge with probability $p$. Under this dynamics, the number of active sites just fluctuates about a certain value. When these wandering boundaries meet, however, either two clusters coalesce with a low probability $p$ through the $101 \to 111$ channel or one cluster vanishes with a high probability $1-p$ through the $010 \to 000$ channel. This behaviour is reminiscent of the long time behaviour of the one-species lattice gas where $k$ particles coalesce into $\ell$ particles, $k{A} \to \ell{A}$ with $k > \ell$, which can be mapped into the problem of the reunion of $k$ random walkers bounded to move in a limited region \cite{mario}. The average density of active sites for very small $p$ is then expected to decay very slowly towards zero as $\rho(t) \sim 1/\sqrt{pt}$ or, equivalently, the extinction time $\tau_{\rm ext}(L,p) \sim L^{2}/p$, with prefactors depending on the sizes and spatial distribution of clusters $11{\cdots}1$ on the initial configuration. Since we sample only a very tiny fraction of all possible initial configurations (for $L=400$ and $\rho(0)=0.3$, the number of possible configurations under periodic boundary conditions is $\sim 1.4 \times 10^{102}$), these prefactors vary widely from one simulation to the other, even if we average over several thousands of realizations. This became evident as the data for $\tau_{\rm ext}(L,p)$ turned out to be overdispersed, {\it i.e.\/}, with sample variance exceeding the mean. Numerical evidence for the scaling of $\tau_{\rm ext}$ is given in Figure~\ref{fig:tau}. We believe that the deviation from the expected exponents is due both to finite-size effects and to the prefactor issue. While it is a little bothering to obtain such non-expected figures, at least we got the correct functional form $\tau_{\rm ext}(L,p) \sim L^{a}/p^{b}$ with $a >1$ and $b \approx 1$, corroborating our qualitative analysis. Notice that $a \simeq z_{\rm DP} \simeq 1.581$ ({\it cf.\/} Section~\ref{indices}), but since this region of small $p$ is very far from the critical region there is no reason to confuse the two values---the figures are similar by accident and by the numerical limitations and issues mentioned before.

The unusual bump in $\rho_{L}(p)$ visible in Figure~\ref{fig:rho} near $p \simeq 0.57$ does not have a simple explanation. We thought at first that it could be signaling a first order phase transition, but further simulation data convinced us that the phase transition is continuous. Notice that the non-monotonic behaviour of $\rho(p)$ for $p \gtrsim 0.57$ is also observed in the mean-field approximations of orders $n=3$ and $4$. We found, however, that the closely related PCA $p150$--$q200$ may provide a clue to this bump. PCA $p150$--$q200$ differs from PCA $p182$--$q200$ in the transition $101 \to 111$, that lacks in the former. It means that in PCA $p150$--$q200$, clusters $11{\cdots}1$ cannot coalesce upon encountering. Preliminary numerical simulations indicate that this PCA suffers an extinction-survival phase transition by $p \simeq 0.572$, in the vicinity of the bump seen in Figure~\ref{fig:rho}. The density of active cells in PCA $p150$--$q200$ is smooth at $p > p_{c}$, without features. A possible scenario for the dynamics around $p \simeq 0.572$ in PCA $p182$--$q200$ is then that the effect of sites surviving and spreading together with that of clusters $11{\cdots}1$ coalescing increase the density of active sites by producing larger clusters that both diminish the erosion from the boundaries (by diminishing the total number of boundaries) and create more stable segments $111$. As $p$ increases from $p \simeq 0.572$, boundaries fluctuate more widely, eventually leading to the annihilation of previously existing clusters that cannot recover easily, leading to a decrease in the density of active sites and to the non-monotonic behaviour seen in Figure~\ref{fig:rho}. Increasing $p$ makes coalescence to become more probable and single active cells in triplets $010$ more stable, leading to a resume of the increase in the density of active sites.

We shall henceforth focus on the transition around $p \simeq 0.48$ only.

\subsection{Critical indices}
\label{indices}

We determined the critical point $p_{c}$ through an examination of the stationary density of active sites $\rho_{L}(p)$ near $p \simeq 0.48$ for different lattice sizes and also by perusal of the scaling relation
\begin{equation}
\rho_{L}(t) \sim t^{-\beta/\nu_{\|}}\Phi(\varepsilon t^{1/\nu_{\|}},\, t^{\nu_{\perp}/\nu_{\|}}/L),
\end{equation}
with $\varepsilon = p-p_{c} \geq 0$ \cite{haye}. For a very large system, this scaling relation assumes the form $\rho(t) \sim t^{-\beta/\nu_{\|}}\Phi(\varepsilon t^{1/\nu_{\|}})$, with $\Phi(x \ll 1) \sim {\rm constant}$ and $\Phi(x \gg 1) \sim x^{\beta}$. The first approach gives $p_{c}$ and $\beta$ more directly but is harder computationally, due to the very long simulation runtimes to attain stationarity close to the critical point. The second approach is more computer-friendly and usually more precise but requires first the simultaneous determination of $p_{c}$ and $\delta = \beta/\nu_{\|}$ and then the determination of $\nu_{\|}$. The third exponent $\nu_{\perp}$ (or, equivalently, $z = \nu_{\|}/\nu_{\perp}$) is obtained by finite-size scaling.

\begin{figure}
\centering
\includegraphics[viewport= 73 72 511 765, scale=0.35, angle=-90]{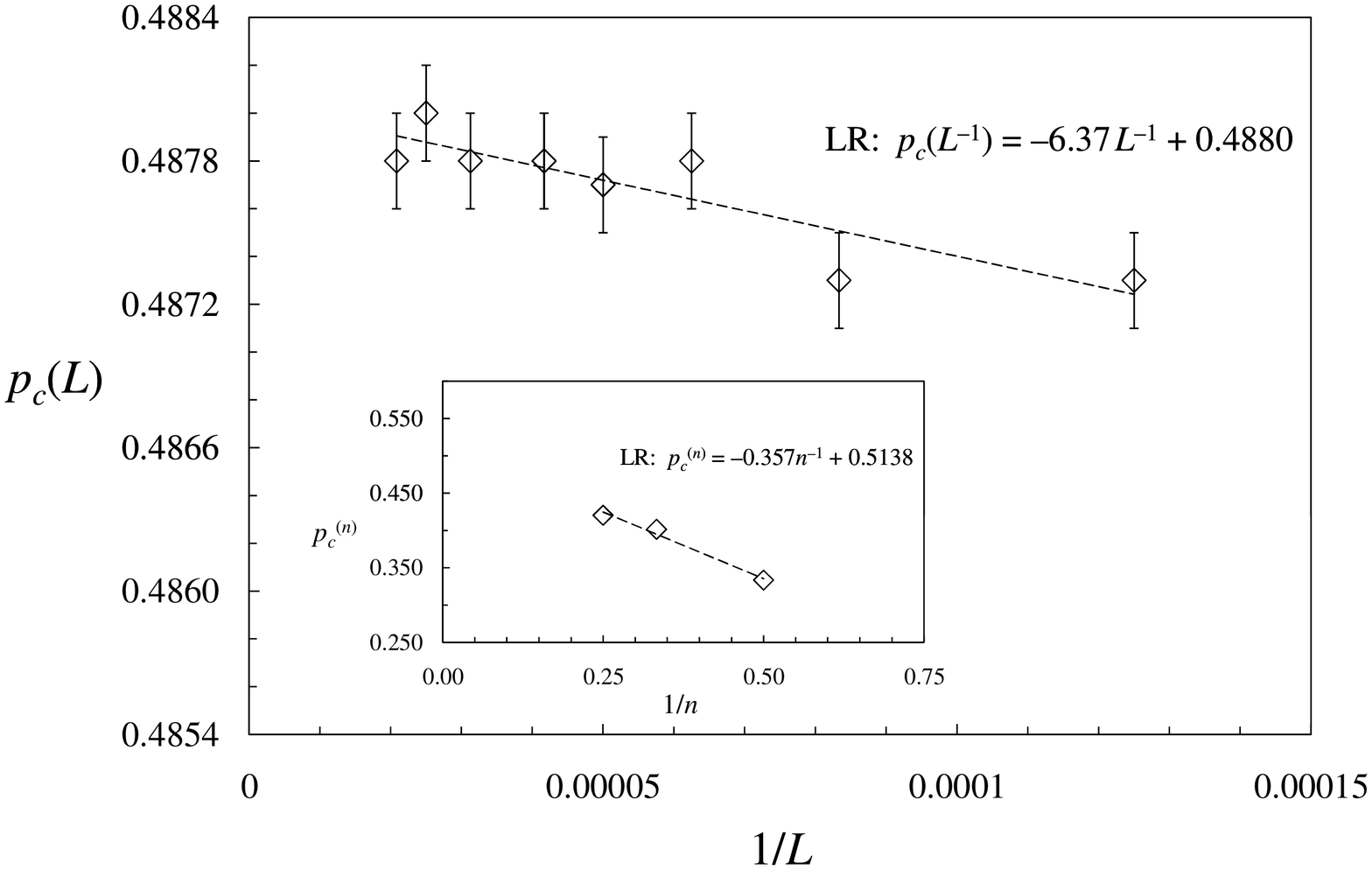}
\caption{\label{fig:pxl}\small Critical points $p_{c}(L)$ for several lattice lenghts $8000 \leq L \leq 48000$. A linear fit gives $p_{c}(L^{-1}) = (-6.37 \pm 1.40)L^{-1}+(0.4880 \pm 0.0001)$ with an $R^{2} = 0.776$. The inset exhibits data from the mean-field approximations of orders $n=2$, $3$, and $4$.}
\end{figure}

Figure~\ref{fig:pxl} shows the estimated critical points $p_{c}(L)$ obtained directly from the stationary density profiles $\rho_{L}(p)$. A linear regression (LR) fit to the data gives $p_{c}(\infty) = 0.4880(1)$ (with a puny $R^{2}=0.776$, though). The inset in the figure exhibits the data from the mean-field approximation of orders $n=2$, 3, and 4 (remember that the $n=1$ and $n=2$ approximations give the same results). The extrapolated $p_{c}^{(\infty)} = 0.5138$ from an LR fit to these data ($R^{2}=0.984$) overshoots the more reliable value obtained from the Monte Carlo simulations and is given here just to illustrate the slow convergence of the mean-field approximation with~$n$.

As we have already mentioned, the determination of $\beta$ through a log-log plot of $\rho(\varepsilon) \sim \varepsilon^{\beta}$ close to the critical point is a poor choice of method. A better option is to plot the time-dependent profile $\rho_{L}(t)$ close to $p_{c}$ for some large $L$. On the critical point, $\rho(t) \sim t^{-\delta}$ and we can estimate $p_{c}$ and $\delta$ simultaneously by plotting $\log_{b}[\rho_{L}(t/b)/\rho_{L}(t)]$ against $1/t$ for some small $b$. Our data for $\rho_{L}(t)$ appear in Figure~5a and represent averages over $1000$ realizations of the dynamics, all with initial density $\rho_{L}(0)=0.2$ in a lattice of $L=16000$ sites. From these data we estimated $p_{c}=0.48810(5)$. The data for the instantaneous values of $\delta$ using $b=10$ appear in Figure~5b. Our best estimate for this exponent comes from the curve with $p=0.4881$ and is given by $\delta = 0.17(1)$.

\begin{figure}
\centering
\begin{tabular}{cc}
\includegraphics[viewport=68 89 515 758, scale=0.25, angle=-90]{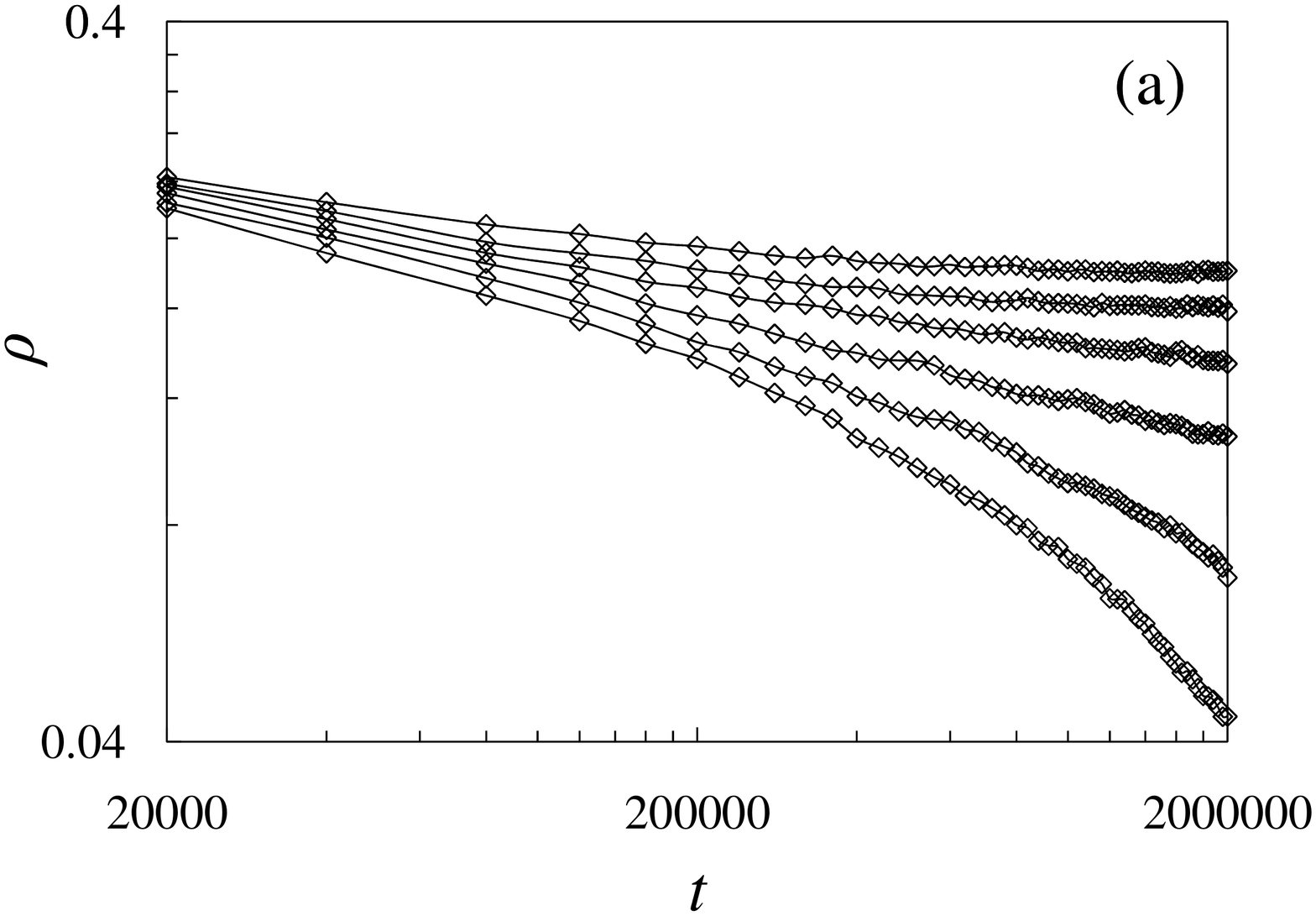} &
\includegraphics[viewport=65 75 531 763, scale=0.25, angle=-90]{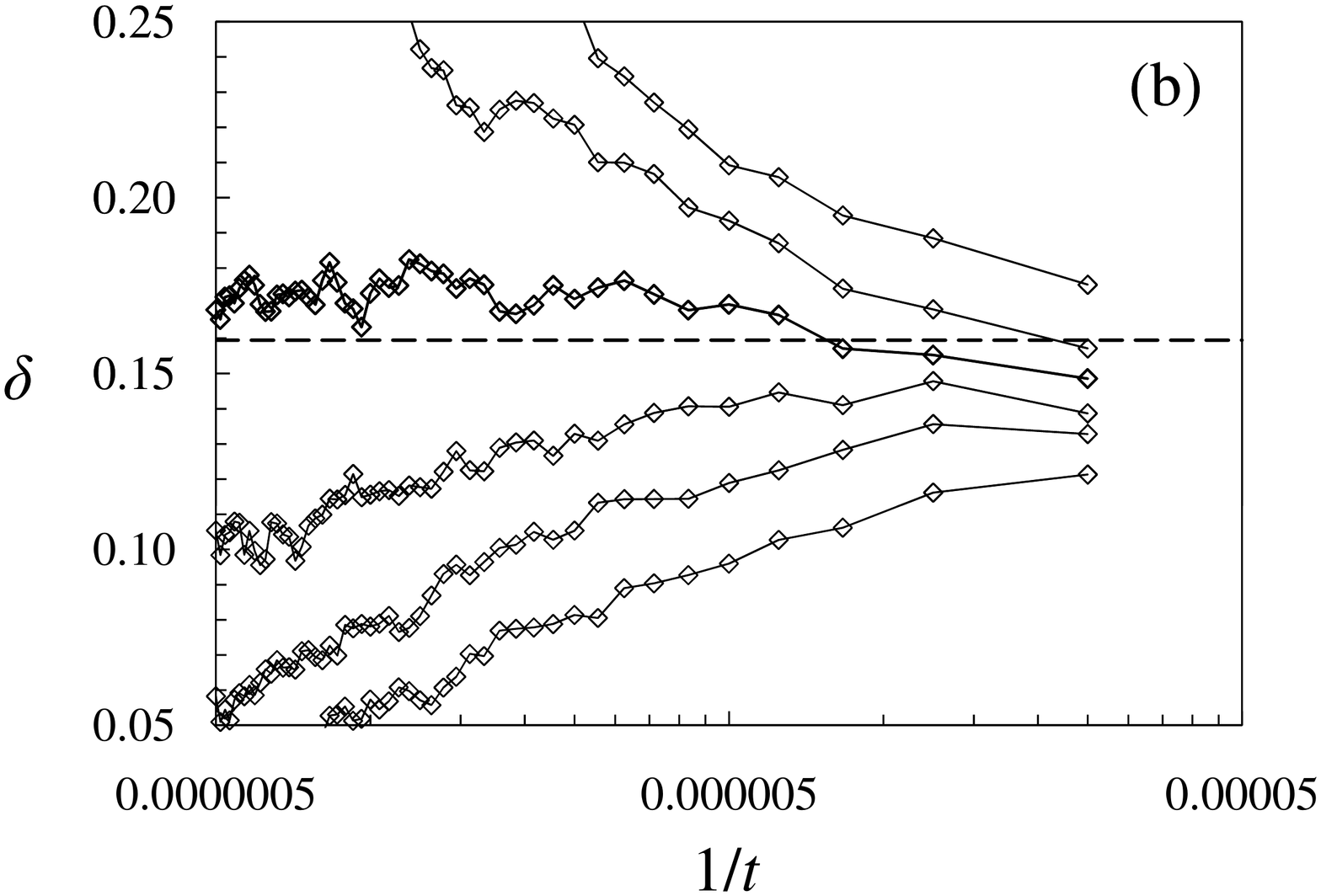}
\end{tabular}
\caption{\label{fig:log}\small (a)~Logarithmic plot of $\rho_{L}(t)$ at fixed $\varepsilon$ with $L=16000$ averaged over $1000$ samples. From the lowermost curve upwards, $p=0.4879$, $0.4880$, $0.4881$, $0.4882$, $0.4883$, and $0.4884$. From this dataset we estimated $p_{c}=0.48810(5)$. (b)~Instantaneous values of $\delta$ obtained from the data depicted in Figure~5a. In panel (b) the uppermost curve corresponds to $p=0.4879$. From the curve with $p=0.4881$ we estimated $\delta = 0.17(1)$. The dashed line corresponds to the best value available for $\delta_{\rm DP}$.}
\end{figure}

\begin{figure}
\centering
\begin{tabular}{cc}
\includegraphics[viewport=68 89 515 758, scale=0.25, angle=-90]{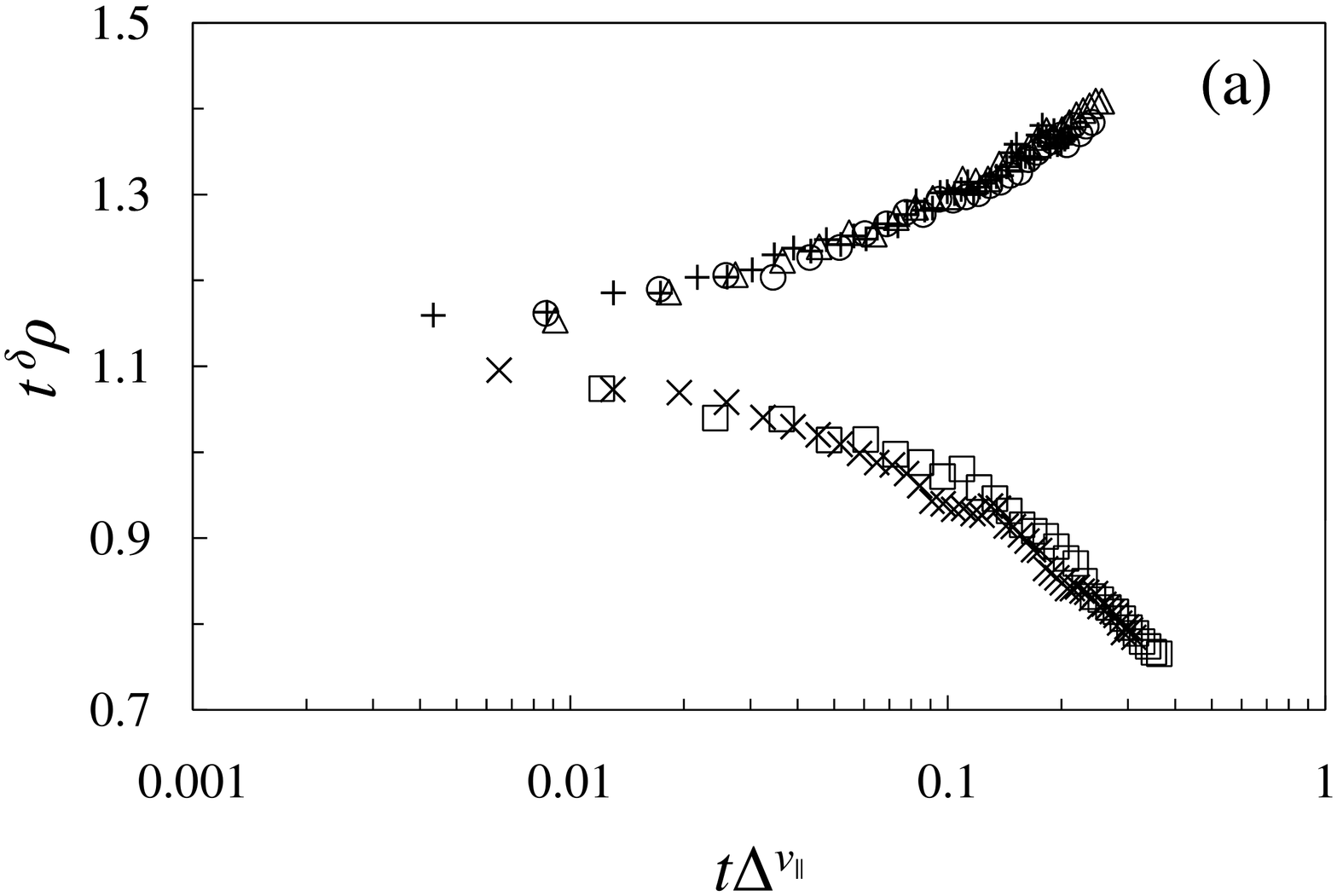} &
\includegraphics[viewport=68 91 515 759, scale=0.25, angle=-90]{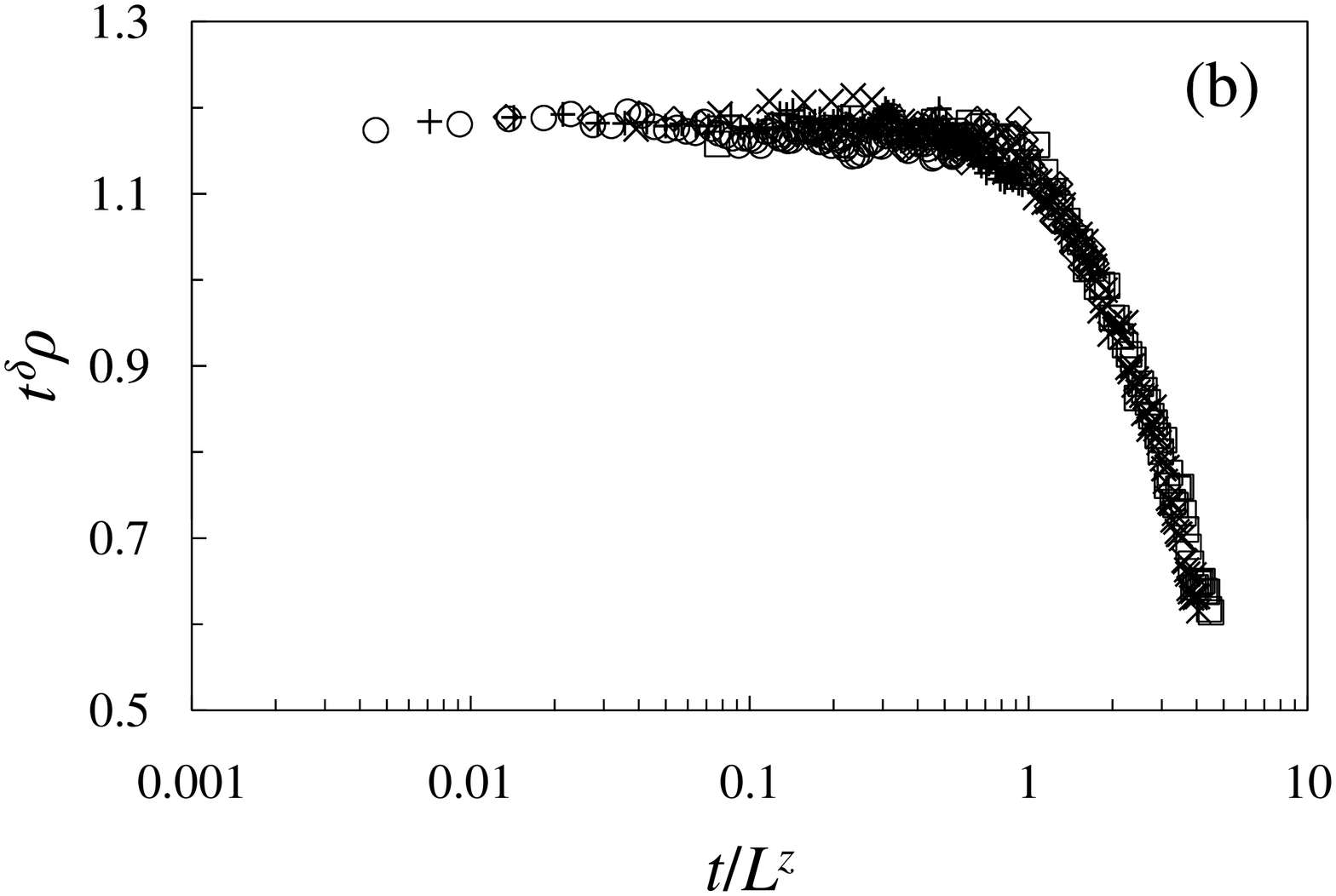}
\end{tabular}
\caption{\label{fig:zet}\small (a)~Data collapse of the scaled time-dependent density profiles for $\varepsilon = \pm 0.0001$, $\pm 0.0002$, and $0.0003$. The upper (lower) branches correspond to $p > p_{c}$ ($p < p_{c}$). The best data collapse was obtained with $p_{c}=0.48812$, $\delta=0.16$, and $\nu_{\|}=1.70$. (b)~Finite-size data collapse of the scaled time-dependent density profiles on the critical point $p_{c} = 0.4881$ for $2000 \leq L \leq 16000$. Best data collapse was obtained with $\delta=0.165$ and $z=1.55$.}
\end{figure}

The exponent $\nu_{\|}$ can be obtained by plotting $t^{\delta}\rho_{L}(t)$ versus $t \varepsilon^{\nu_{\|}}$ and tuning $\nu_{\|}$ to achieve data collapse with different $\varepsilon$. Similarly, by plotting $t^{\delta}\rho_{L}(t)$ versus $t/L^{z}$ at $p_{c}$ for different $L$ and tuning $z$ until data collapse we can obtain $z = \nu_{\|}/\nu_{\perp}$. The collapsed curves for our data appear in Figure~\ref{fig:zet}. Figure~6a was obtained with $p=0.48812$, $\delta=0.16$, and $\nu_{\|}=1.7$. We can comfortably estimate $\nu_{\|}$ only within $\pm 0.05$, so that our best estimate for this exponent is $\nu_{\|} = 1.70(5)$. Combining $\delta$ and $\nu_{\|}$ furnishes $\beta = \delta \nu_{\|} = 0.29(2)$. Data collapse in Figure~6b was obtained for data taken at $p = 0.4881$ with $\delta = 0.165$ and $z = 1.55$. As before, we could determine $z$ only within $\pm 0.05$, so that our best estimate for it reads $z = 1.55(5)$. The exponents $\delta$, $\nu_{\|}$, and $z$ together determine the universality class of critical behaviour of the model, other exponents following from well known hyperscaling relations \cite{haye}.

The best values available for the critical exponents of the DP process on the $(1+1)D$ square lattice are $\delta_{\rm DP} = 0.159\,464(6)$, $\nu_{\|{\rm DP}} = 1.733\,847(6)$, $\beta_{\rm DP} = 0.276\,486(8)$, and $z_{\rm DP} = 1.580\,745(10)$ \cite{jensen}. Thus, within the error bars our data put the critical behaviour of PCA $p182$--$q200$ in the DP universality class. Our estimation of $\delta$ is somewhat high and limitrophe to the accepted $\delta_{\rm DP}$. We probably underestimated the uncertainty in this quantity. This higher value of $\delta$ impacted the estimation of $\nu_{\|}$ and $z$, which turned out to be smaller than the accepted values, but in these cases still within the error bars. Anyway, we can safely establish the universality class of critical behaviour of PCA $p182$--$q200$ as that of the DP process, as no other universality class has exponents close to the values found for it.

\section{Summary and perspectives}
\label{summary}

We showed that the mixture of CA rules $182$ and $200$ into a PCA suffers an extinction-survival-type phase transition at $p_{c}=0.48810(5)$ in the $(1+1)D$ directed percolation universality class of critical behaviour. Although PCA $p182$--$q200$ has two absorbing configurations, $00{\cdots}0$ and $11{\cdots}1$, this second absorbing configuration is hardly achieved, since neither CA~$182$ nor CA~$200$ have a stationary density of active sites close to $1$ except if the initial configuration is $11{\cdots}1$ itself. We also found that PCA $p182$--$q200$ has some unusual features, notably its slow dynamics for very small $p$, when it approaches the stationary state diffusively with a large relaxation time proportional to $1/p$, and the bump in its density profile.

We would like to delve further into the small $p$ behaviour of PCA $p182$--$q200$. In this regard, it would not be entirely without interest to consider a stochastic reaction-diffusion version of the PCA consisting of the reactions $100 \rightleftharpoons 110$, $001 \rightleftharpoons 011$, and $010 \to 000$, making up a sort of constrained, partially reversible contact process. In this model, clusters $11{\cdots}1$ never coalesce and can only be eroded from the boundaries. It is also related with an RSOS interface model introduced some time ago \cite{aehm}. We guess that, depending on the rate of the process $010 \to 000$, it displays an extinction-survival phase transition. Moreover, simple diffusion $10 \rightleftharpoons 01$ may represent a relevant perturbation for this lattice gas, since it destabilizes $11{\cdots}1$ clusters and provides room for more $010 \to 000$ reactions.

There is plenty of room for exploratory incursions into possibly interesting single-parameter composite PCA. Of particular interest would be to find composite PCA displaying phase transitions in the even branching and annihilating random walk or the directed Ising universality classes of critical behaviour \cite{odor}.


\section*{Acknowledgments}

The authors acknowledge partial financial support from the Conselho Nacional de Desenvolvimento Cient\'{\i}fico e Tecnol\'{o}gico -- CNPq, Brazil.



\begin{thebibliography}{99}

\bibitem{jvn}J. von Neumann, {\it Theory of Self-Reproducing Automata\/}, edited by A. W. Burks (University of Illinois Press, Urbana {\&} London, 1966).

\bibitem{discrete}A. L. Toom, N. B. Vasilyev, O. N. Stavskaya, L. G. Mityushin, G. L. Kurdyumov, and S. A. Pirogov, ``Discrete local Markov systems,'' in: {\it Stochastic Cellular Systems: Ergodicity, Memory, Morphogenesis\/}, edited by R. L. Dobrushin, V. I. Kryukov, and A. L. Toom (Manchester University Press, Manchester, 1990), pp.~1--182.

\bibitem{wolfram}S. Wolfram, {\it Cellular Automata and Complexity: Collected Papers\/} (Addison-Wesley, Reading, 1994).

\bibitem{dkpca}E. Domany and W. Kinzel, ``Equivalence of cellular automata to Ising-models and directed percolation,'' \href{http://dx.doi.org/10.1103/PhysRevLett.53.311}{{\it Phys. Rev. Lett.\/} {\bf 53}, 311--314 (1984)}; W. Kinzel, ``Phase transitions of cellular automata,'' \href{http://dx.doi.org/10.1007/BF01309255}{{\it Z. Phys. B\/} {\bf 58}, 229--244 (1985)}.

\bibitem{grinstein}G. Grinstein, C. Jayaprakash, and Y. He, ``Statistical mechanics of probabilistic cellular automata,'' \href{http://dx.doi.org/10.1103/PhysRevLett.55.2527}{{\it Phys. Rev. Lett.\/} {\bf 55}, 2527--2530 (1985)}.

\bibitem{rujan}P. Ruj\'{a}n, ``Cellular automata and statistical mechanical models,'' \href{http://dx.doi.org/10.1007/BF01009958}{{\it J. Stat. Phys.\/} {\bf 49}, 139--222 (1987)}.

\bibitem{lebowitz}J. L. Lebowitz, C. Maes, and E. R. Speer, ``Statistical mechanics of probabilistic cellular automata,'' \href{http://dx.doi.org/10.1007/BF01015566}{{\it J. Stat. Phys.\/} {\bf 59}, 117--170 (1990)}.

\bibitem{chen}S. Chen and G. D. Doolen, ``Lattice Boltzmann method for fluid flows,'' \href{http://dx.doi.org/10.1146/annurev.fluid.30.1.329}{{\it Annu. Rev. Fluid Mech.\/} {\bf 30}, 329--364 (1998)}.

\bibitem{helbing}D. Helbing, ``Traffic and related self-driven many-particle systems,'' \href{http://dx.doi.org/10.1103/RevModPhys.73.1067}{{\it Rev. Mod. Phys.\/} {\bf 73}, 1067--1141 (1998)}.

\bibitem{droz}B. Chopard and M. Droz, \href{http://dx.doi.org/10.2277/0521673453}{{\it Cellular Automata Modeling of Physical Systems\/}} (Cambridge University Press, Cambridge, 2005).

\bibitem{boccara}F. Bagnoli, N. Boccara, and R. Rechtman, ``Nature of phase transitions in a probabilistic cellular automaton with two absorbing states,'' \href{http://dx.doi.org/10.1103/PhysRevE.63.046116}{{\it Phys. Rev. E\/} {\bf 63}, 046116 (2001)}.

\bibitem{fuks}H. Fuk\'{s}, ``Solution of the density classification problem with two cellular automata rules,'' \href{http://dx.doi.org/10.1103/PhysRevE.55.R2081}{{\it Phys. Rev. E\/} {\bf 55}, R2081--R2084 (1997)}.

\bibitem{noisy}J. R. G. Mendon\c{c}a, ``Sensitivity to noise and ergodicity of an assembly line of cellular automata that classifies density,'' \href{http://dx.doi.org/10.1103/PhysRevE.83.031112}{{\it Phys. Rev. E\/} {\bf 83}, 031112 (2011)}.

\bibitem{tania}T. Tom\'{e}, ``Spreading of damage in the Domany-Kinzel cellular automaton: A mean-field approach,'' \href{http://dx.doi.org/10.1016/0378-4371(94)90139-2}{{\it Physica A\/} {\bf 212}, 99--109 (1994)}.

\bibitem{mario}M. J. de Oliveira, ``Diffusion-limited annihilation and the reunion of bounded walkers,'' \href{http://dx.doi.org/10.1590/S0103-97332000000100012}{{\it Braz. J. Phys.\/} {\bf 30}, 128--132 (2000)}.

\bibitem{haye}H. Hinrichsen, ``Nonequilibrium critical phenomena and phase transitions into absorbing states,'' \href{http://dx.doi.org/10.1080/00018730050198152}{{\it Adv. Phys.\/} {\bf 49}, 815--958 (2000)}.

\bibitem{jensen}I. Jensen, ``Low-density series expansions for directed percolation: I. A new efficient algorithm with applications to the square lattice,'' \href{http://dx.doi.org/10.1088/0305-4470/32/28/304}{{\it J. Phys. A: Math. Gen.\/} {\bf 32}, 5233--5249 (1999)}; M. A. Mu\~{n}oz, R. Dickman, A. Vespignani, and S. Zapperi, ``Avalanche and spreading exponents in systems with absorbing states,'' \href{http://dx.doi.org/10.1103/PhysRevE.59.6175}{{\it Phys. Rev. E\/} {\bf 59}, 6175--6179 (1999)}; J. R. G. Mendon\c{c}a, ``Precise critical exponents for the basic contact process,'' \href{http://dx.doi.org/10.1088/0305-4470/32/44/102}{{\it J. Phys. A: Math. Gen.\/} {\bf 32}, L467--L473 (1999)}.

\bibitem{aehm}U. Alon, M. R. Evans, H. Hinrichsen, and D. Mukamel, ``Roughening transition in a one-dimensional growth process,'' \href{http://dx.doi.org/10.1103/PhysRevLett.76.2746}{{\it Phys. Rev. Lett.\/} {\bf 76}, 2746--2749 (1996)}.

\bibitem{odor}G. \'{O}dor, ``Universality classes in nonequilibrium lattice systems,'' \href{http://dx.doi.org/10.1103/RevModPhys.76.663}{{\it Rev. Mod. Phys.\/} {\bf 76}, 663--724 (2004)}.

\end{thebibliography}
\end{document}